\documentclass[preprintnumbers, amsmath, amssymb, showpacs,twocolumn, superscriptaddress, 
nofootinbib]{revtex4}

\usepackage{graphicx}
\usepackage{epsfig}
\usepackage{bm}
\usepackage{amssymb}
\usepackage{float}
\usepackage{amsmath}
\usepackage{dcolumn}
\usepackage{cancel}
\usepackage[colorlinks]{hyperref}
\usepackage[usenames,dvipsnames]{color}
\hypersetup{
     breaklinks=true,
    pdfstartview={FitH},    
    colorlinks=true,       
    linkcolor=blue,          
    citecolor=red,        
    filecolor=magenta,      
    urlcolor=blue,           
    anchorcolor=green,      
    linktocpage=true
}

\begin{document}

\title{Correspondence of cosmology from non-extensive 
thermodynamics with fluids of generalized equation of state   
 }

 \author{Shin'ichi Nojiri}
\email{nojiri@gravity.phys.nagoya-u.ac.jp}
\affiliation{Department of Physics, Nagoya University, Nagoya 464-8602, Japan}
\affiliation{Kobayashi-Maskawa Institute for the Origin of Particles and the
Universe, Nagoya University, Nagoya 464-8602, Japan}

\author{Sergei D. Odintsov}
\email{odintsov@ieec.uab.es}
\affiliation{Institut de Ciencies de lEspai (IEEC-CSIC), 
Campus UAB, Carrer de Can Magrans, s/n 
08193 Cerdanyola del Valles, Barcelona, Spain}
\affiliation{Instituci\'{o} Catalana de Recerca i Estudis Avan\c{c}ats
(ICREA), Passeig Llu\'{i}s Companys, 23 08010 Barcelona, Spain}
\affiliation{Department of General and Theoretical Physics, Eurasian 
International Center for 
Theoretical Physics, Eurasian National University,  Nur-Sultan, Kazakhstan}
\affiliation{Int.Lab. Theor. Cosmology, Tomsk State University of
Control Systems and Radioelectronics, 634050 Tomsk, Russia}

\author{Emmanuel N. Saridakis}
\email{msaridak@phys.uoa.gr}
\affiliation{Department of Physics, National Technical University of Athens, Zografou
Campus GR 157 73, Athens, Greece}
\affiliation{National Observatory of Athens, Lofos Nymfon, 11852 Athens, 
Greece}
\affiliation{Department of Astronomy, School of Physical Sciences, University of Science 
and Technology of China, Hefei 230026, P.R. China}
\affiliation
{Department of General and Theoretical Physics, Eurasian 
International Center for 
Theoretical Physics, Eurasian National University,  Nur-Sultan, Kazakhstan}

\author{R. Myrzakulov}\email{rmyrzakulov@gmail.com}
\affiliation{Department of General and Theoretical Physics, Eurasian 
International Center for 
Theoretical Physics, Eurasian National University,  Nur-Sultan, Kazakhstan}

\begin{abstract}  
We show that there is a correspondence between cosmology from non-extensive thermodynamics 
and cosmology with fluids of redefined and generalized equation of state. We first 
establish the correspondence in the case of basic non-extensive thermodynamics, and then 
we proceed by investigating the more consistent case, from the quantum field theoretical 
point of view, of varying exponent, namely depending on the scale. The obtained duality 
provides a way of explaining the complicated phenomenological forms of the effective fluid 
equation-of-state parameters that are being broadly used in the literature, since their 
microphysical origin may indeed lie in the non-extensive thermodynamics of spacetime.
Finally, concerning the cosmological behavior, we show that at late times the effective 
fluid may drive the universe acceleration even in the absence of an explicit cosmological 
constant, and even if the initial fluid is the standard dust matter one. Similarly, at 
early times we obtain an effective cosmological constant which is enhanced through 
screening, and hence it can drive a a successful inflation without spoiling the correct 
late-time acceleration. 

\end{abstract}


\maketitle

\section{Introduction \label{Sec1}}

In the recent years there has appeared an increasing amount of cosmological data, from 
early, intermediate and late times, whose successful explanation may imply the need of 
modifying our current knowledge. The modification of gravity (for general reviews see
\cite{Capozziello:2011et,Nojiri:2010wj,Nojiri:2017ncd}), 
is one of the two ways that is being followed in order to obtain the extra degrees of 
freedom required, with the second one being the introduction of the inflaton  
\cite{Bartolo:2004if} and/or dark energy fields/fluids \cite{Peebles:2002gy,Cai:2009zp}.
 
Amongst the  various approaches to modified gravity one can find the interesting one that 
is based on the connection between 
gravity and thermodynamics \cite{Jacobson:1995ab,Padmanabhan:2003gd,Padmanabhan:2009vy}.
In such a consideration one applies the first law of thermodynamics in the universe 
horizon and results to the Friedmann 
equations, and it proves to be true in   various classes of modified gravity 
 \cite{Cai:2005ra,Akbar:2006kj,Cai:2006rs,Akbar:2006er,Paranjape:2006ca,
Sheykhi:2007zp,Jamil:2009eb,
Cai:2009ph,Wang:2009zv,Jamil:2010di,Gim:2014nba,Fan:2014ala,DAgostino:2019wko}.
In this thermodynamical approach one needs to apply the entropy expression of the 
involved theory. Although the standard Boltzmann-Gibbs additive entropy is the one that 
it is usually used, the fact that in the case of non-additive systems, such as large-scale 
gravitational ones, such an entropy should be  
generalized to the non-extensive Tsallis entropy 
\cite{Tsallis:1987eu,Lyra:1998wz,Wilk:1999dr}, led to the thermodynamic considerations of 
gravity based on the latter 
\cite{Tsallis:2012js,Komatsu:2013qia,Nunes:2014jra,Lymperis:2018iuz,Saridakis:2018unr,
Sheykhi:2018dpn,Artymowski:2018pyg,Abreu:2017hiy,Jawad:2018frc,Zadeh:2018wub,
daSilva:2018ehn}. Indeed,   parametrizing the 
non-extensivity  by a new exponent $\delta$, for which  the value 1 corresponds to 
the standard entropy, one may obtain modified Friedmann equations,
which may lead to various (old and new) cosmological scenarios compatible with 
observations \cite{Lymperis:2018iuz}. 
This approach has been further generalized to the non-extensive 
thermodynamics in which the involved exponent presents a running behavior 
\cite{Nojiri:2019skr}, which  is typical for quantum field theory when 
renormalization group is incorporated. Cosmology    from extended entropy with 
varying exponent is able to provide a successful description of the universe 
evolution at both late and early times.

On the other hand, it is known that one can describe the early and late time universe 
evolution by considering perfect fluids  with negative pressure, satisfying a
barotropic equation of state 
\cite{Li:2011sd,Barrow:1994nx,Tsagas:1998jm,
HipolitoRicaldi:2009je,Gorini:2005nw,Kremer:2003vs,
Carturan:2002si,Brevik:2017juz,Buchert:2001sa,Hwang:2001fb,Elizalde:2009gx, Grande:2011xf,
Cruz:2011zza,Lima:2012mu,Oikonomou:2017mlk,Capozziello:2006dj,Basilakos:2009wi,
Koivisto:2015qua, Paliathanasis:2015arj,
Elizalde:2017dmu,
Brevik:2016kuy,
Balakin:2012ee,Brevik:2018azs} or presenting  generalized equation-of-state parameters 
\cite{Nojiri:2005sr,Nojiri:2006zh,Elizalde:2005ju} (for a review see 
\cite{Brevik:2017msy}). In such scenarios the fluids are considered to arise effectively, 
without the need of explaining their microphysical origin.
 
In the present work we show that there is a correspondence between cosmology from 
non-extensive thermodynamics and cosmology with fluids of redefined and generalized 
equation of state. This form of duality may be important in providing the unknown  
microphysical origin of the effective fluids that are used broadly in cosmology. 
Additionally, we present how different epochs of the universe evolution can be 
successfully described.

The plan of the work is as follows: In Section \ref{Sec3} we demonstrate the 
correspondence between cosmology from non-extensive thermodynamics  with fluids of 
generalized equation of state, starting from the former and resulting to the latter, as 
well as 
starting from the latter and resulting to the former. In Section \ref{Sec4} we present 
the 
same analysis in the extended case where the non-extensive exponent presents a varying 
behavior, depending on the scale, as it is the case of a  quantum field theory when 
renormalization group is incorporated. Moreover, we apply the scenario at both early- and 
late-time universe. We close this work with our conclusions in Section \ref{Conclusions}.

\section{Correspondence of non-extensive 
thermodynamics with fluids of redefined equation of state    }
 \label{Sec3}

In this section we show how cosmology with fluids with redefined equation-of-state 
parameters can be obtained from non-extensive thermodynamics and vice versa.

\subsection{Fluids of redefined equation of state through non-extensive 
thermodynamics }

We start by 
briefly reviewing how modified cosmology can arise through the application of 
non-extensive, Tsallis thermodynamics. We apply a homogeneous and isotropic flat
Friedmann-Robertson-Walker (FRW) geometry with metric
 \begin{equation}
\label{FRW}
ds^2 = - dt^2 + a(t)^2 \sum_{i=1,2,3} \left( dx^i \right)^2 \, ,
\end{equation}
with  $a(t)$ the scale factor.
Moreover, we consider the universe to be filled with 
a perfect fluid, with energy density and pressure  $\rho$ and $p$ respectively. Such a 
system may be considered as a thermodynamical system bounded by the the apparent horizon  
$r_H$ defined   using the 
Hubble
rate $H\equiv \dot a/a$ as   \cite{Cai:2005ra,Cai:2008gw}
\begin{equation}
\label{Tslls1}
r_H=\frac{1}{H} \, .
\end{equation}
The energy $dE$ going outward through the sphere of radius  $r_H$
in a time interval $dt$ is given by \cite{Cai:2005ra}
\begin{equation}
\label{Tslls2}
dQ = - dE = \frac{4\pi}{3} r_H^3 \dot\rho dt = \frac{4\pi}{3H^3} \dot\rho dt \, 
.
\end{equation} 
By using the standard conservation law
\begin{equation}
\label{Tslls3}
0 = \dot \rho + 3 H \left( \rho + p \right) \, ,
\end{equation}
one can rewrite Eq.~(\ref{Tslls2}) as  
\begin{equation}
\label{Tslls4}
dQ = \frac{4\pi}{H^2} \left( \rho + p \right) dt \, .
\end{equation}  

The next step is to introduce the entropy through
\begin{equation}
\label{Tslls6B}
TdS = dQ \, ,
\end{equation}
with  the Hawking temperature $T$ given by
\cite{Cai:2005ra,Cai:2008gw}
\begin{equation}
\label{Tslls6}
T = \frac{1}{2\pi r_H} = \frac{H}{2\pi}\, .
\end{equation}
If the usual entropy relation $S = \frac{A}{4G}$ (with $A=4\pi r_H^2$ the horizon area) 
is used then one can immediately see that (\ref{Tslls4}),(\ref{Tslls6B}),(\ref{Tslls6}) 
give the usual Friedmann equations \cite{Cai:2005ra,Cai:2008gw}. However, if instead of 
the usual entropy ones applies the generalized, non-extensive one, namely Tsallis entropy 
 \cite{Tsallis:1987eu,Tsallis:2012js}, which is the correct one to be used in 
non-extensive systems such as large-scale 
gravitational ones and it is given by  
\begin{equation}
\label{Tslls9}
S = \frac{A_0}{4 G} \left(\frac{A}{A_0} \right)^\delta\, , 
\end{equation}
with $A_0 \equiv \frac{4\pi}{H_0^2} $, $H_0$ a   constant introduced for dimensional 
reasons  and $\delta$ the non-extensive exponent, then 
(\ref{Tslls4}),(\ref{Tslls6B}),(\ref{Tslls6}) lead to 
\cite{Saridakis:2018unr,Lymperis:2018iuz}
 \begin{equation}
\label{Tslls10}
\delta \left( \frac{H_0^2}{H^2} \right)^{\delta -1} \dot H
=  - 4\pi G \left( \rho + p \right) \, .
\end{equation}
 Thus, inserting (\ref{Tslls3}) and
integrating  we obtain  
\begin{equation}
\label{Tslls11}
\frac{\delta}{2 - \delta} H_0^2\left( \frac{H^2}{H_0^2} \right)^{2 - \delta}
= \frac{8\pi G}{3} \rho + \frac{\Lambda}{3} \, ,
\end{equation}
where  the cosmological constant $\Lambda$ appears as a constant of   
integration. Equations (\ref{Tslls10}),(\ref{Tslls11}) are the two Friedmann equations of 
modified cosmology through non-extensive horizon thermodynamics 
\cite{Saridakis:2018unr,Lymperis:2018iuz}. Note that in the standard case of  
$\delta=1$ they become the standard Friedmann equations.
  
In the present work we show that the above equations can lead to usual cosmology however 
with  the incorporation of fluids of redefined equation of state. For simplicity we 
consider the case $\Lambda=0$ and we re-write  Eq.~(\ref{Tslls11})  as 
\begin{equation}
\label{Tslls12}
\frac{3}{8 \pi G} H^2 = \tilde \rho \equiv
C_0 \rho^{\frac{1}{2-\delta}} \, ,  
\end{equation}
where 
\begin{equation}
\label{Tslls12b} 
C_0 \equiv \frac{3}{8\pi G}\left( 4\pi \right)^{\frac{1 - \delta}{2 - \delta}}
\left[ \frac{8\pi G \left( 2 - \delta \right) }{3\delta}
\right]^{\frac{1}{2-\delta}}  \, .
\end{equation}
In the case where the initial perfect fluid has a constant
equation of state (EoS) parameter $w$,  namely $p=w\rho$,   $\rho$ behaves as
$\rho \propto a^{ - 3 \left( 1 + w \right)}$ and therefore $\tilde \rho$
behaves as $\tilde\rho \propto a^{ - \frac{3 \left( 1 + w \right)}{2 - 
\delta}}$, which implies that the EoS parameter is effectively changed as
\begin{equation}
\label{Tslls13}
w_\mathrm{eff} = - 1 + \frac{1 + w}{2 - \delta} \, .
\end{equation}
Using the conservation law (\ref{Tslls3}) we acquire
\begin{equation}
\label{NTslls1}
\dot{\tilde\rho} = - 3H \frac{1}{2-\delta}C_0 \rho^{- 
\frac{1-\delta}{2-\delta}}
\left( \rho + p \right) \, ,
\end{equation}
and thus we may define the effective pressure $\tilde p$ as
\begin{equation}
\label{NTslls2}
\tilde p \equiv \frac{1}{2-\delta}C_0 \rho^{- \frac{1-\delta}{2-\delta}}
\left( \rho + p \right) - \tilde \rho \, ,
\end{equation}
in order for  $\tilde\rho$ and $\tilde p$ to satisfy the conservation law
$0=\dot{\tilde\rho} + 3 H \left( \tilde\rho + \tilde p \right)$. 
Hence, (\ref{NTslls2}) can be finally written as
\begin{equation}
\label{NTslls3}
\tilde p = \left( - 1 + \frac{1 + w}{2 - \delta} \right) \tilde\rho\, ,
\end{equation}
which is consistent with (\ref{Tslls13}) as expected. As we observe, through the 
application of non-extensive thermodynamics in a cosmological framework we were able to 
obtain standard cosmology but with fluids of redefined equation-of-state parameter.

\subsection{Non-extensive 
thermodynamics from fluids of generalized equation of state   }

Let us now proceed by considering  an inhomogeneous fluid with generalized  EoS  of the 
form
\cite{Nojiri:2005sr}
\begin{equation}
\label{NTsllsB1}
p = f \left( \rho, H, \dot H, \ddot H, \cdots \right) \, ,
\end{equation}
where $f$ is a function of $\rho$, $H$, $\dot H$, $\ddot H$, etc.  In this case 
Eqs.~(\ref{Tslls12}) and (\ref{NTslls2}) give
\begin{eqnarray}
\label{NTsllsB2}
&&\!\!\!\!\!\!\! 
\tilde p = \tilde f \left( \tilde\rho, H, \dot H, \ddot H, \cdots \right)\nonumber\\
&&\!\!\!\!\!\!\! 
\equiv \frac{C_0 }{2\! -\! \delta}\! 
\left[ C_0^{\delta-2} \tilde\rho^{2-\delta}
+ f \! \left( C_0^{\delta-2} \tilde\rho^{2-\delta}, H, \dot H, \ddot H, \cdots 
\right) \! \right]  \rho^{- \frac{1-\delta}{2-\delta}}  \nonumber\\
&&
  - \tilde \rho  \, .
\end{eqnarray}
Therefore, the non-extensivity of modified cosmology can be absorbed into 
the redefinition of the
general EoS of the cosmological fluid. 

As an example we may consider a fluid with  the   
EoS
\begin{equation}
\label{NTsllsB3}
p=g \left( \rho \right) + f_0 H^{\beta} \, ,
\end{equation}
where  $f_0$ and $\beta$ are constants and $g$ is a function of the energy 
density $\rho$.
Then Eq.~(\ref{NTsllsB2}) leads to  the following EoS:
\begin{equation}
\label{NTsllsB4}
\tilde p = \frac{C_0}{2\! -\! \delta}\, \rho^{- \frac{1-\delta}{2-\delta}}
\left[ C_0^{\delta-2} \tilde\rho^{2-\delta}
+ g \left( C_0^{\delta-2} \tilde\rho^{2-\delta} \right) + f_0 H^{\beta} 
\right]
  - \tilde \rho  \, .
\end{equation}
In standard cosmology with a perfect fluid with a constant EoS parameter $w$ (with 
$w\neq-1$) it is known that 
\begin{equation}
\label{NTslls4}
H= \frac{2}{3(w+1)t}\,,
\end{equation}
which implies that $ a= a_0 t^{\frac{2}{3(w+1)}}$, and thus $\rho = \rho_0 a^{ - 3 \left( 
1 
+ w \right)}$ gives 
\begin{equation}
\label{NTslls4b}
\rho(t)= \rho_0 a_0^{ - 3 \left( 1 + w \right)} t^{-2}\, ,
\end{equation}
with  $a_0$ and $\rho_0$ being constants.
Then the energy $Q$ in the comoving volume $V=V_0 a^3 = V_0 a^3 
t^{\frac{2}{w+1}}$,
with a constant $V_0$, is given
by
\begin{equation}
\label{NTslls5}
E= -Q = \rho V = \rho_0 V_0 a^{-3w} = \rho_0 V_0 a_0^{-3w} t^{- \frac{2w}{w + 
1}}\, .
\end{equation}
Furthermore, applying Hawking temperature  (\ref{Tslls6})  using
(\ref{NTslls4}) we find
\begin{equation}
\label{NTslls5B}
T = \frac{1}{3(w+1)\pi t} \, .
\end{equation}
Hence, inserting the above into the thermodynamical relation (\ref{Tslls6B}) gives
\begin{equation}
\label{NTslls6}
dS = 12\pi \rho_0 V_0 a_0^{-3w} t^{- \frac{2w}{w + 1}} dt \,  ,
\end{equation}
which for $w\neq 1$  leads to 
\begin{equation}
\label{NTslls6b}
S  = \frac{12\pi(w+1)}{1-w}  \rho_0 V_0 a_0^{-3w} t^{- \frac{2w}{w + 1}} \, .
\end{equation}
Hence we resulted to a relation $S\propto V^{-w}$, i.e. we recovered non-extensive 
thermodynamics. This is the inverse procedure of the previous subsection, and thus it 
completes the correspondence     of fluids with redefined equation-of-state 
parameter and  non-extensive thermodynamics.

\section{ Extensions
\label{Sec4}}

In this section we extend the analysis of the previous section in the case where the 
non-extensive exponent  $\delta$ of Tsallis entropy 
(\ref{Tslls9}) depends on the energy scale, i.e. it presents a running behavior
\cite{Nojiri:2019skr}. This arises from the fact that 
the entropy corresponds to physical 
degrees of freedom, nevertheless the renormalization of a quantum theory implies 
that the degrees of freedom depend on the scale.
In case of gravity,  if the spacetime  fluctuates  at   high energy 
scales then the degrees of freedom may increase, however if gravity becomes 
topological then the
degrees
of freedom may decrease, which  shows that in general  the exponent $\delta$ may depend 
on the scale.

For   cosmological considerations the energy scale may be given by the 
Hubble scale $H$, and thus $\delta$ may depend on $H$ \cite{Nojiri:2019skr}.
We   parametrize the dependence by using $x\equiv \frac{H_1^2}{H^2} $, with  $H_1$
a parameter that has dimensions identical to $H$.
Then by following the procedure of the previous section, instead of 
(\ref{Tslls10}) one finds \cite{Nojiri:2019skr}
\begin{equation}
\label{Tslls16}
\left\{ \delta\! 
+ \! \left[ \frac{H_1^2}{H^2} \ln \left( \frac{H_1^2}{H^2} \right) \right]
\delta'
\right\} \left( \frac{H_1^2}{H^2} \right)^{\delta -1} \dot H
= - 4\pi G \left( \rho + p \right) \, ,
\end{equation}
where $\delta'(x)\equiv d \delta(x)/d x$.
Integrating (\ref{Tslls16}) and using (\ref{Tslls3}) one finds
\begin{equation}
\label{Tslls17}
\left. - H_1^2 \left\{ x^{\delta(x) - 2} + 2 \int^x dx x^{\delta(x) -3} 
\right\}
\right|_{x=\frac{H_1^2}{H^2}}= \frac{8\pi G}{3} \rho + \frac{\Lambda}{3} \, .
\end{equation}
Equations (\ref{Tslls16}),(\ref{Tslls17}) are the generalized Friedmann equations that 
arise from non-extensive thermodynamics of varying exponent.

We  consider the same general scenario of \cite{Nojiri:2019skr}, namely we choose
\begin{equation}
\label{Tslls17Bdelta}
\delta(x) = \frac{ \ln \left[c \left( x^{3-n} + \alpha(x) b_2 x^{2-n}
+ b_1 b_2^2 x^{1-n}\right) \right] }{\ln x}\, ,  
\end{equation}
with 
\begin{equation}
\label{Tslls17Bdeltab} 
\alpha (x)\equiv \frac{n (3 - n)}{(1-n)^2} + \frac{n^2}{(1+n)(2-n)} b_1 b_2^2
x^{-n-1}\, ,
\end{equation}
 and where $c \equiv \left\{ \frac{3 - n}{(1-n)^2} + \frac{b_1}{1+n} 
\right\}^{-1}b_2^{n-2} $, with
 $n$, $b_1$, $b_2$ the model parameters.
For the scenario (\ref{Tslls17Bdelta}), Eq.~(\ref{Tslls17}) becomes
\begin{equation}
\label{Tslls17Ca}
\left. - f(x) \right|_{x=\frac{H_1^2}{H^2}} H_1^2
= \frac{8\pi G}{3} \rho + \frac{\Lambda}{3}\, ,  
\end{equation}
with 
 \begin{align}
\label{Tslls17Cabb}
f(x) \equiv c & \left[ \left( \frac{3 - n}{1-n}\right) x^{1-n} 
 -\left( \frac{2 - n}{n}\right) b_2 \alpha(x) x^{-n}\right.\nonumber\\
&\ \left. -\left( \frac{1-n}{1+n}\right) b_1 b_2^2 x^{-n-1} \right] \, .
\end{align}
Finally, we mention that for the choices   $n=2$ and $b_1=b_2=0$ we obtain $\delta(x)=1$, 
i.e. standard thermodynamics, and in this case
Eq.~(\ref{Tslls17Ca}) becomes the standard Friedmann equation
\begin{equation}
\label{Tslls8}
H^2 = \frac{8\pi G}{3} \rho + \frac{\Lambda}{3} \, .
\end{equation}

Let us proceed by showing that the modified cosmology from non-extensive thermodynamics 
with varying exponent can be re-written as standard cosmology but with fluids with 
generalized equation-of-state parameter. In particular,  Eq.~(\ref{Tslls17}) can 
be rewritten in the standard form  
\begin{equation}
\label{NTslls7}
\frac{3}{8 \pi G} H^2 = \tilde \rho \,,
\end{equation}
  if we define
\begin{equation}
\label{NTslls7b} 
\tilde \rho\equiv \frac{3 H_1^2}{8 \pi G
\mathcal{F}^{-1} \left( \frac{8\pi G}{3} \rho + \frac{\Lambda}{3} \right)} \, ,
\end{equation}
where $\mathcal{F}^{-1}$ is the inverse function of  
\begin{equation}
\label{NTslls8}
\mathcal{F}(x) \equiv - H_1^2 \left\{ x^{\delta(x) - 2}
+ 2 \int^x dx x^{\delta(x) -3} \right\} \, ,
\end{equation}
or equivalently 
\begin{equation}
\label{NTslls7B}
\rho = \frac{3}{8 \pi G} \left[ - \frac{\Lambda}{3}
+ \mathcal{F} \left(\frac{3 H_1^2} {8 \pi G \tilde\rho} \right) \right] \, .
\end{equation}
We can now define   the effective pressure as
\begin{equation}
\label{NTslls9}
\tilde p \equiv - \frac{\dot{\tilde\rho}}{3H} - \tilde \rho \, ,
\end{equation}
in order for   $\tilde\rho$ and $\tilde p$ to satisfy the conservation law.
Hence, the effective fluid acquires a generalized equation-of-state parameter of the form 
\begin{equation}
\label{NTslls10B}
w_\mathrm{eff} \equiv \frac{\tilde p}{\tilde \rho} 
= - 1 - \frac{8\pi G}{3}\rho
( 1 \! + \! w ) 
\left[ \ln \mathcal{F}^{-1} \left( \frac{8\pi G}{3} \rho
+ \frac{\Lambda}{3} \right) \right]'  ,
\end{equation}
with $w=p/\rho$ the EoS parameter of the initial fluid and with primes denoting 
derivative 
with respect to the argument.

We now proceed by performing the inverse of the above approach. We consider   a fluid 
with a generalized equation of state of the form (\ref{NTsllsB1}), and in particular 
with 
 \begin{eqnarray}
\label{NTslls10}
&&
\!\!\!\!\!\!\!\!\!\!\!\!\!
\tilde p = 
\tilde  f \left( \tilde \rho, H, \dot H, \ddot H, \cdots \right)  
\equiv
- \frac{ \left( 8 \pi G \tilde\rho \right)^2}{  
9 H_1^2 \mathcal{F}' \left(\frac{3 H_1^2} {8 \pi G \tilde\rho} \right)
}\nonumber\\
&&
\!\!
\cdot
   \left\{ 
    f \left( \frac{3}{8 \pi G} \left[ - \frac{\Lambda}{3}
+ \mathcal{F} \left(\frac{3 H_1^2} {8 \pi G \tilde\rho} \right) \right],
H, \dot H, \ddot H, \cdots \right)
\right.
\nonumber\\
&&\left. \ \
+ \frac{3}{8 \pi G}
\left[ - \frac{\Lambda}{3}
+ \mathcal{F} \left(\frac{3 H_1^2} {8 \pi G \tilde\rho} \right) \right] \right\}
- \tilde\rho \, .
\end{eqnarray}
Hence, one can immediately see that this fluid has an effective EoS parameter of the form 
(\ref{NTslls10B}).

In summary, the result of the previous section holds also in the extended case where the 
exponent $\delta$ presents a varying behavior, namely there is a correspondence between 
non-extensive thermodynamics and fluids with generalized EoS.  This   duality is one of 
the main results of the present work, and it provides a way of explaining the complicated 
phenomenological forms of the effective fluid EoS parameters that are being broadly used 
in the literature. Namely, their microphysical origin may lie in the non-extensive 
thermodynamics of spacetime.

The compelling feature of the scenario at hand is that one can obtain interesting 
cosmological phenomenology even if the initial fluid is the standard dust matter. For 
instance, one can see that for the case $w\geq0$, if 
 $4\pi G
\left( 1 + w \right)\rho 
\left[ \ln \mathcal{F}^{-1} \left( \frac{8\pi G}{3} \rho
+ \frac{\Lambda}{3} \right) \right]' <1$ then $w_\mathrm{eff}<-1/3$ and thus  
the expansion of the universe is accelerated, while if 
$\left[ \ln \mathcal{F}^{-1} \left( \frac{8\pi G}{3} \rho
+ \frac{\Lambda}{3} \right) \right]' <0$ then  $w_\mathrm{eff}<-1$  and the phantom 
regime is realized. We mention that these conditions may be obtained even in the case 
where the explicit cosmological constant is set to zero, namely for $\Lambda=0$, which is 
an
advantage that reveals the capabilities of the scenario at hand, since in this case the 
universe acceleration results purely from the non-extensivity, or equivalently from the 
generalized fluid equation of state.

We close this section by focusing on the early universe, namely we apply  
  (\ref{Tslls17Bdelta}) at the early times where $H\gg H_1$ and therefore $x\ll 1$.
  We study separately case I where $b_1\neq 0$  and case II where  $b_1=0$ and $b_2\neq0$.
  For case I the Friedmann equation becomes 
\begin{equation}
\label{Tslls17G}
c b_1 \left(\frac{1-n}{1+n}\right) \left( \frac{H_1^2}{H^2} \right)^{-n} H^2
= \frac{8\pi G}{3} \rho + \frac{\Lambda}{3} \, ,
\end{equation}
while for case II we have
\begin{equation}
\label{Tslls17H}
  (2-n) b_2^{n-1} \left( \frac{H_1^2}{H^2} \right)^{1-n} H^2
= \frac{8\pi G}{3} \rho + \frac{\Lambda}{3}\, .
\end{equation}
Hence, concerning the  energy density of the effective fluid, for case I we find
\begin{equation}
\label{NTslls10BB}
\tilde\rho =  \frac{3}{8\pi G} \left\{ \frac{H_1^{2n}}{cb_1} \left( 
\frac{1+n}{1-n} \right)
\left( \frac{8\pi G}{3} \rho + \frac{\Lambda}{3} \right) 
\right\}^{\frac{1}{1+n}} \, ,  
\end{equation}
while for case II
\begin{equation}
\label{NTslls11}
\tilde\rho =  \frac{3}{8\pi G} \left\{ \frac{H_1^{2\left(n-1\right)}}{b_2^{n-1}
\left( 2 - n \right) }
\left( \frac{8\pi G}{3} \rho + \frac{\Lambda}{3} \right) \right\}^{\frac{1}{n}} 
\,  .
\end{equation}
 Therefore, using (\ref{NTslls9}) we can calculate the pressure of the effective fluid, 
which for case I reads as  
\begin{eqnarray}
\label{NTslls12}
&&\tilde p = \frac{( 1 + w )\rho}{1+n} \left[ \frac{H_1^{2n}}{cb_1} \left( 
\frac{1+n}{1-n} 
\right)
\right]^{\frac{1}{1+n}}
\left( \frac{8\pi G}{3} \rho + \frac{\Lambda}{3} \right)^{-\frac{n}{1+n}} 
\nonumber\\
&& \ \ \  \ \ 
  -  \frac{3}{8\pi G} \left[ \frac{H_1^{2n}}{cb_1} \left( \frac{1+n}{1-n} 
\right)
\left( \frac{8\pi G}{3} \rho + \frac{\Lambda}{3} \right) 
\right]^{\frac{1}{1+n}} \, ,
\end{eqnarray}
while for case II as
\begin{eqnarray}
\label{NTslls13}
&&
\tilde p = \frac{( 1 + w )\rho}{n} \left[ \frac{H_1^{2\left(n-1\right)}}{b_2^{n-1}
\left( 2 - n \right) } \right]^{\frac{1}{n}}
\left( \frac{8\pi G}{3} \rho + \frac{\Lambda}{3} \right)^{\frac{1-n}{n}}  \nonumber\\
&& \ \ \  \ \ \
  - \frac{3}{8\pi G} \left[ \frac{H_1^{2\left(n-1\right)}}{b_2^{n-1}
\left( 2 - n \right) }
\left( \frac{8\pi G}{3} \rho + \frac{\Lambda}{3} \right) \right]^{\frac{1}{n}} 
\, .
\end{eqnarray}
 Since we are in the early universe we can neglect the standard matter fluid, and thus 
the above expressions become
\begin{equation}
\label{NTslls16}
\tilde p = - \tilde\rho
= - \frac{3}{8\pi G} \left[ \frac{H_1^{2n}}{cb_1} \left( \frac{1+n}{1-n} 
\right)
\left( \frac{\Lambda}{3} \right) \right]^{\frac{1}{1+n}} \, ,
\end{equation}
for case I, and
\begin{equation}
\label{NTslls17}
\tilde p = - \tilde\rho
= - \frac{3}{8\pi G} \left[ \frac{H_1^{2\left(n-1\right)}}{b_2^{n-1}
\left( 2 - n \right) } \left( \frac{\Lambda}{3} \right) \right]^{\frac{1}{n}} 
\, ,
\end{equation}
for case II.
Hence, as we can see, we can define the effective cosmological constant as
\begin{equation}
\label{Tslls17M2}
\Lambda_\mathrm{eff}\equiv 3 \left[ \frac{(1+n) \Lambda H_1^{2n}}{3(1-n) cb_1 }
\right]^{\frac{1}{n+1}}
\end{equation}
for case I, and as
\begin{equation}
\label{Tslls17HH2}
\Lambda_\mathrm{eff}\equiv 3\left[ \frac{ \Lambda H_1^{2(n-1)}}{3 (2-n) 
b_2^{n-1}}
\right]^{\frac{1}{n}}
\end{equation}
for case II.
Eqs.~(\ref{Tslls17M2}) and (\ref{Tslls17HH2}) indicate that the cosmological
constant is effectively screened, which is an advantage since 
it allows to obtain an inflation realization even if $\Lambda$ is small enough in order 
to be consistent with the late-time universe acceleration.

At this point we should add a comment on the relation of the present scenario 
with the model of inflation in double-screen entropic cosmology
\cite{Cai:2010zw,Cai:2010kp}. In such model one applies the entropic 
gravity approach, that arises from holographic considerations, and obtains 
extra 
terms in the Friedmann equations that depend on higher powers of the energy 
density. These terms can then drive a successful inflation in which the  
holographic statistics on the outer screen may lead to a sizable value of the 
non-linearity parameter \cite{Cai:2010zw,Cai:2010kp}. Indeed, having in mind 
the above analysis, one can see that the extra terms  that depend on higher 
powers of the energy density correspond effectively to fluids with generalized 
equation of state, and thus falling inside the general class of the scenarios 
of the present work.

In summary, the scenario at hand may simultaneously describe   both   the inflation era
(where $H^2 \sim \left( 10^{24}\, \mathrm{eV} \right)^2$), as well as the late-time 
acceleration epoch  (where $H^2 \sim \left( 10^{-33}\, \mathrm{eV} \right)^2$), even if 
the only fluid that is present in the universe is the standard dark matter one. This is 
a significant advantage and it is one of the main results of this work.

\section{Conclusions}
 \label{Conclusions}

In the present work we showed that there is a correspondence between cosmology from 
non-extensive thermodynamics and cosmology with fluids of redefined and generalized 
equation of state. In particular, it is well known that one can obtain modified cosmology 
through the application of thermodynamics in the universe horizon,  and this approach has 
been recently extended through the use of non-extensive entropy. On the other hand it is 
also known that one can describe the universe phenomenology through the use of effective 
fluids with generalized equation-of-state parameter of unknown microphysical origin. 

We first established the above correspondence in the case of basic 
non-extensive thermodynamics, showing first that cosmology from non-extensive 
thermodynamics can result to effective fluids with redefined EoS parameters, and then
completing the picture by showing that cosmology with fluids of redefined EoS results to 
non-extensive thermodynamics.

We proceeded by investigating the extended case of non-extensive 
thermodynamics of varying exponent, namely when the exponent depends on the scale, which 
is a   more consistent case quantum field theoretically. As we showed, we also 
established the above correspondence, nevertheless the involved fluids acquire a 
generalized EoS parameter and not just a redefined one. 
This duality is one of the main results of the present work, since it provides a way 
of explaining the complicated phenomenological forms of the effective fluid EoS parameters 
that are being broadly used in the literature. In particular, their microphysical origin 
may lie exactly in the non-extensive thermodynamics of spacetime.

Concerning the cosmological behavior, we showed that at late times the effective fluid may 
drive the universe acceleration even in the absence of an explicit cosmological constant, 
and even if the initial fluid is the standard dust matter one. Similarly, at early times 
we obtain an effective cosmological constant which is enhanced through screening, and 
hence it can drive a a successful inflation without spoiling the correct late-time 
acceleration.

In summary, the established correspondence between cosmology from non-extensive 
thermodynamics and cosmology with fluids with generalized equation of state provides a 
theoretical justification of the latter, and hence the cosmological application deserves 
further investigation.

\begin{acknowledgments}

This work is partly supported   by MEXT KAKENHI Grant-in-Aid for 
Scientific Research on Innovative Areas g Cosmic Acceleration h No. 15H05890 
(S.N.) 
and the JSPS Grant-in-Aid for Scientific Research (C) No. 18K03615 (S.N.), 
and by MINECO (Spain), FIS2016-76363-P, and 
by project 2017 SGR247 (AGAUR, Catalonia) (S.D.O). The work was carried out with 
the financial support of the Ministry of Education 
and Science of the Republic of Kazakhstan, Grant No 0118RK00935. S.D.O. 
acknowledges the hospitality of  Eurasian National University where this work 
was done.
\end{acknowledgments}

\end{document}